\documentstyle[editedbook,epsfig,psfig,epsf,times,mathptm,bm]{mq}

\def\etal{{\em et al.} }

\def\cm2{cm$^2$ }
\def\se1{s$^{-1}$ }

\def\bm#1{\mbox{\boldmath{$#1$}}}

\begin{opening}
\title{The Theory of Relativistic Jet Formation in Galactic Sources: 
Towards a Unified Model}
\author{David L. Meier$^1$}
\institute{$^1$ Jet Propulsion Laboratory, California Institute of Technology, Pasadena, CA, USA.}
\end{opening}

\runningtitle{Relativistic Jet Formation}
\runningauthor{Meier}

\begin{document}
\vspace{-0.5cm}
\begin{abstract}
{\small 
I review recent progress in the theory of relativistic jet production. 
The presently favored mechanism is an electrodynamic one, in which 
charged plasma is accelerated by electric fields that are generated by a
rotating magnetic field.  The most pressing issues of current interest
are understanding what factors control the jet power, its speed, and
its degree of collimation, and how these properties determine the type
of jet observed and its effect on its environment.\\
Recent observations of microquasars, pulsars, gamma-ray bursts (GRBs)
and core-collapse supernovae indicate that jets play an important, and in some
cases possibly dominant, role in each of these.  The presence of a
jet in all cases may provide an important clue to how these sources
may be related. Based on these observations, and on recent theoretical
investigations, I propose an evolutionary scheme that attempts to unify
all of these relativistic galactic jet sources. I also discuss several
important issues that must be resolved before this (or another scheme)
can be adopted.}
\end{abstract}

\section{Introduction: An Expanded Definition of ``Microquasar''}
\label{sec:intro}
The standard definition of ``microquasar'' --- the object that is 
the subject of this meeting --- is a binary black hole candidate X-ray
source that possesses a relativistic jet at least part of the time in
its life cycle.    For the purpose of this talk, and to illustrate the
unified model proposed, I will use an expanded definition of microquasar,
stated as follows.  A microquasar is any galactic, stellar-mass source
that produces a jet whose velocity is a significant fraction ($> 10\%$)
of the speed of light $c$.  This definition includes the following sources:

\begin{enumerate}
\item Classical microquasars (GRS 1915+105, GRO J1655-40, GX 339-4,
etc.):  These produce jets with $v_{jet} > 0.6 - 0.95 c$ 
($\Gamma_{jet} \equiv [ 1 - v_{jet}^{2} / c^{2} ]^{-1/2} > 1.25 - 3$).
 
\item Gamma-ray bursts (GRBs):  Believed to be ``microquasars in
formation'' and produce jets with $\Gamma_{jet} \sim 100 - 300$ that 
point toward us.

\item SS433-type objects:  Observed properties suggest a super-Eddington 
accreting, magnetized neutron star.  Jets have speed $v_{jet} \sim 0.25 c$.

\item Isolated pulsars (Crab, Vela, etc.):  Jets now have been detected by
Chandra in these objects and have speeds $v_{jet} \sim 0.5 c$.

\item Core-collapse supernovae:  There is growing evidence that supernovae 
also produce jets with a power comparable to the explosion itself.
Expected speeds are $v_{jet} \sim 0.25 - 0.5 c$ --- the escape speed from the
new proto-neutron star.
\end{enumerate}

The inclusion of core-collapse supernovae (SN) above is the key to the 
unified model.  In the mid 1990s it was discovered that such supernovae 
emit polarized light in the optical band \cite{wang01,leo01}, caused by 
electron scattering by an asymmetrically-expanding explosion.  The 
variation of polarization properties with time and
with different SN types gives important clues to its nature.
In a given SN, the degree of polarization $\Pi$ increases with time
but the polarization direction remains constant in time and wavelength,
indicating that the asymmetry is global and maintains a fixed direction.  
For Type IIa SN (ones with a large hydrogen-rich envelope), the 
$\Pi \sim 1\%$, indicating only a 2:1 or less asymmetry.  For Type IIb SN
(ones that have lost their hydrogen envelope prior to the explosion,
leaving only the helium envelope), $\Pi$ is higher ($\sim$2\%),
indicating a 2.5:1 axial ratio.  For SN Type Ib/Ic (ones that have
lost most or all of their envelope, leaving a compact blue Wolf-Rayet
star that then explodes), $\Pi$ is quite high ($4-7$\%),
indicating a 3:1 axial ratio or better.  Clearly, the deeper one sees
into the explosion, the more elongated the exploding object appears.
The observers have concluded that core-collapse
SN have a global prolate shape that appears to be associated
with the central engine producing the explosion.  A jet, with energy
comparable to that of the explosion itself, significantly alters the
shape of the envelope, creating the elongated, polarized central source.
In the paper below we will show that all five of the above ``microquasars''
are intimately related, both in the physical origins of the jet itself
and in their astrophysical origins as well.

\section{Basic Principles of Magnetohydrodynamic Jet Production}
\label{sec:basic_principles}
The basic principles of MHD jet production have been described elsewhere 
\cite{meier01a}. The reader is referred to that paper for a more detailed
description, and more comprehensive figures, than in the short review
given below.

\subsection{The Jet Engine Itself:  Launching of the Jet Outflow}
\label{sec:jet_launching}

\subsubsection{Jet Production in Accreting Systems and Pulsars} 
\label{sec:generic_mhd_jets} 
Several mechanisms for producing bipolar outflows have been suggested
(explosions in the center of a rotationally-flattened cloud,
radiation-pressure-driven outflows from a disk, etc.), but none of
these is able to produce outflows approaching the highly-relativistic
speeds observed in the fastest jet sources.  The currently-favored
mechanism is a magnetohydrodynamic one, somewhat similar to terrestrial 
accelerators of particle
beams.  Indeed, electromagnetic acceleration of relativistic pulsar
winds has been a leading model for these objects since the 1960s.
MHD (or electrodynamic) jet production was first suggested in 1976 
\cite{bland76,love76} and has been applied model to rotating black holes 
\cite{bz77} (BZ) and to magnetized accretion disks \cite{bp82} (BP).  
This mechanism has now been simulated \cite{su85,kud99,nak01} and is 
sometimes called the ``sweeping pinch'' mechanism.

The most important ingredient in the MHD mechanism is a magnetic field
that is anchored in a rotating object and extends to large distances
where the rotational speed of the field is considerably slower.
Plasma trapped in the magnetic field lines is subject to the Lorentz
(${\bm J} \times {\bm B}$) force, which, under conditions of high 
conductivity (the MHD assumption), splits into two vector components:  
a magnetic pressure gradient ($- {\bm \nabla} B^{2} / 8 \pi$) and a 
magnetic tension (${\bm B} \cdot {\bm \nabla} {\bm B} / 4 \pi$). 
Differential rotation between the inner and outer regions winds up the
field, creating a strong toroidal component ($B_{\phi}$ in cylindrical 
[$R$, $Z$, $\phi$] coordinates).  The magnetic pressure gradient up the rotation axis
($-d B_{\phi}^{2} / d Z$) accelerates plasma up and out of the system while the
magnetic tension or ``hoop'' stress ($-B_{\phi}^{2} / R$) pinches and collimates the
outflow into a jet along the rotation axis.

This basic configuration of differential rotation and twisted magnetic
field accelerating a collimated wind can be achieved in all objects
identified in Sect. \ref{sec:intro}. For classical
microquasars and GRBs, the field will be anchored in
the accreting plasma, which may lie in a rotating disk (BP)
and/or may be trapped in the rotating spacetime of the spinning
central black hole itself (BZ).  In the case of
SS433-type objects, isolated pulsars, and core-collapse SN
the rotating field is anchored in the pulsar (or proto-pulsar). 
In SS433 and core-collapse SN the source of the accelerated 
plasma is, once again, accretion, but in isolated pulsars it is believed
to be particles created in spark gaps by the high ($10^{12}$ G) field.

\subsubsection{The Case of Kerr Black Holes:  Direct and Indirect Magnetic Coupling}
\label{sec:blackhole_mhd_jets}
The jet-production mechanism envisioned by BZ generally
involved direct magnetic coupling of the accelerated plasma to the 
rotating horizon.  That is, magnetic field lines thread the
horizon, and angular momentum is transferred along those field lines to
the external plasma via magnetic tension.  However, another, indirect, 
coupling is possible.  This mechanism, suggested by \cite{pc90} (PC) and recently 
simulated by us \cite{koide02}, has the same effect as the BZ mechanism
(extraction of angular momentum from the rotating black hole by the
magnetic field), but the field lines do not have to thread the horizon
itself.  Instead, they are anchored in the accreting plasma.
When this plasma sinks into the ergosphere near the black hole ($R <
2~G M / c^{2}$), frame dragging causes the plasma to rotate with respect to
the exterior, twisting up the field lines in a
manner similar to the situation when the field is anchored in a disk
or pulsar.  (This occurs even if the accreting plasma has
no angular momentum with respect to the rotating spacetime.) 
The twisted field lines then have two effects:

\begin{enumerate}
\item Electromagnetic power is ejected along the rotation axis in the form
of a torsional Alfven wave.  Eventually the
output Poynting flux power should be dissipated in the production and
acceleration of particles and a fast jet.

\item The back-reaction of the magnetic field accelerates the ergospheric 
plasma (in which it is anchored) to relativistic speeds {\em against} the rotation 
of the black hole.  The counter-rotating ergospheric plasma now formally has 
negative angular momentum and negative energy (negative mass); that is,
it has given up more than its rest mass in energy to the external
environment.  It is on orbits that must intersect the black hole horizon,
and, when it does, the mass of the black hole
decreases by a value equal to that negative energy.
\end{enumerate}

This process is the magnetic equivalent of the Penrose process, but
instead of extracting black hole rotational energy by particle scattering,
the energy is extracted by scattering of an Alfven wave off the ergospheric
plasma particles.  Determining whether the BZ or PC process occurs in 
certain systems is an important question for future study.

\subsubsection{Output Engine Power:  Thick Accretion Flows Make Better Jets}
The output Poynting flux power of the rotating magnetic field depends on
the strength of the poloidal component of the field 
$B_{p0} = (B_{R0}^{2}+B_{Z0}^{2})^{1/2}$, its angular speed 
$\Omega_{0}$, and the size of the rotating region $R_{0}$.  In the
non-relativistic magnetized disk theory of BP the jet power is given by
\begin{equation}
L_{jet}  =  B_{p0}^{2} R_{0}^{3} \Omega_{0}
\label{eq:nonrel_jet_power}
\end{equation}
In the relativistic theory of rotating black holes (BZ) the output power 
is 
\begin{equation}
L_{jet}  =  f  B_{p0}^{2} R_{0}^{4} \Omega_{0}^{2} / (4 c)
\label{eq:rel_jet_power}
\end{equation}
The geometric factor $f$ is of order unity in BZ 
and in several early papers that use these formulae;  
however, others \cite{ga97} have argued for a value $f \sim 1/8$, which reduces the estimated
power by almost an order of magnitude.  Eq.  (\ref{eq:rel_jet_power}) occurs in a variety
of relativistic cases, not just in BZ.  PC's jets are the equivalent 
of the BP disk process, but in a black hole ergosphere; they also find   
an output power with a similar form and $f \sim 1$. 
A similar expression is obtained for pulsar winds if one takes
the output power to be $L_{jet} = \Gamma_{jet} \dot{M}_{jet} c^{2}$, where 
$\dot{M}_{jet}$ is the mass outflow rate, and uses the general pulsar result that 
$\Gamma_{jet} \approx \sigma \equiv B_{p0}^{2} R_{0}^{4} \Omega_{0}^{2}  / ( 4 \dot{M}_{jet} c^{3})$ 
\cite{c89}, the magnetization parameter.
Eq.  (\ref{eq:rel_jet_power}), therefore, generally is used to estimate the power
in relativistic cases (most of those considered here), while eq.  
(\ref{eq:nonrel_jet_power})
is used in non-relativistic situations, such as jets from protostars or
planetary nebulae.  

Eq.  (\ref{eq:rel_jet_power}) has interesting implications for
the output power for accretion disks of various thicknesses.  It has been argued 
\cite{livio99} that the poloidal magnetic field in an accretion disk
is only a fraction of the azimuthal field:  $B_{p0} \sim (H_{0}/R_{0}) B_{\phi 0}$, 
where $H_{0}$ is the disk half-height at radius $R_{0}$ and $B_{\phi 0}$ is the azimuthal 
field that is responsible for the angular momentum transport in the accretion disk
at $R_{0}$. Substituting into eq.  (\ref{eq:rel_jet_power}) we obtain
\begin{equation}
L_{jet}  =  f  H_{0}^{2} B_{\phi 0}^{2} R_{0}^{2} \Omega_{0}^{2} / (4 c)
\label{eq:rel_jet_power_w_height}
\end{equation}
Therefore, thick accretion flows with $H_{0} \sim R_{0}$, such as ADAFs, are 
expected to have much more powerful jets than thin accretion flows and 
$H_{0} << R_{0}$, such as standard accretion disks. We have used this
property \cite{meier01b} to explain the observed presence of jets in the 
microquasar low/hard state and the absence of jets in the high/soft state 
\cite{Fender99}. Recent numerical results \cite{hb02}
which perform 3-dimensional global simulations of the magneto-rotational
instability, show a jet being launched, but only from the inner portion
of the accretion disk where the flow is geometrically thick and the
poloidal magnetic field is substantial.

\subsection{Formation of the Jet:  Acceleration and Collimation}
\label{sec:jet_acc_collim}

\subsubsection{Slow Acceleration and Collimation is Probably the Norm}
\label{sec:slow_acc_normal}
Because the dynamical time scale is of order 0.1 ms or less in these
objects, a steady state is set up fairly quickly in jet ejection events
that last even only a few seconds.  In a steady state, the wind 
accelerates as it expands vertically away from the rotator.  A jet is 
not fully formed until its speed exceeds the local wave propagation
speed, {\it i.e.}, the total Alfven speed 
$V_{A} = [(B_{R}^{2}+B_{Z}^{2}+B_{\phi}^{2}) / (4 \pi \rho)]^{1/2}$, where
$\rho$ is the mass density in the outflowing material.  The place where this
occurs, often called the Alfven point or Alfven surface, generally is
well above the rotating object producing the accelerating torsional
Alfven wave.  Analytic \cite{bp82,li92} and numerical \cite{kras99} studies 
of this steady state show that the outflow is rather broad at the base, and 
it slowly focuses as it is accelerated. At a height $Z_{A} > 10 R_{0}$ 
above the disk, the total Alfven speed is exceeded, the flow is focused into
a narrow cylindrical or conical flow, and little more acceleration and
collimation takes place.  The terminal jet speed $v_{jet}$ is of order 
$V_{A}(Z_{A})$, and this speed is usually of order the escape speed
from the central rotator $V_{esc}(R_{0})$.

There now is observational evidence that, in at least some {\em 
extragalactic} systems, the steady-state picture of slow acceleration and
collimation is correct.  Very high resolution VLBA radio images of the
M87 jet \cite{junor99} show a broad opening angle at the base that
narrows to only a few degrees after a few hundred Schwarzschild radii.
In addition, it has been argued \cite{sm01} that most quasar
jets must be broad at the base:  they lack soft, Comptonized
and relativistically-boosted X-ray emission that would be expected from
a narrow, relativistic jet flow near the black hole.  For 
microquasars, a similar result still would imply a collimation region
much too small to resolve with current telescopes --- $100 r_{S} \sim 10^{8}$ cm
or 2 nano-arcseconds at a distance of 3 kpc.

\subsubsection{Rapid Acceleration and Collimation May be Possible for 
Strong Fields} \label{sec:fast_acc_possible}
We have found that when the MHD luminosity of the jet exceeds a critical value
\begin{equation}
L_{crit}  =  E_{escape} / \tau_{free-fall}  =  4\pi \rho_{0} R_{0}^{2} (G M / R_{0})^{3/2}
\label{eq:magswitch_jet_power}
\end{equation}
rapid acceleration of the plasma can take place \cite{meier97}. 
In this case, magnetic forces will exceed gravitational and centrifugal
forces and the field will dynamically pinch the plasma above the rotator.
Angular momentum conservation will cause this pinched material to spin
very rapidly, and if this material then re-connects to the rotator below, strong
shear in the re-connected region will create an even stronger toroidal
magnetic field that rapidly accelerates and collimates the outflow. The
resulting jet is very fast, possibly significantly exceeding the escape
speed $V_{esc}(R_{0})$, and highly-collimated within a few $R_{0}$.  We have
called this process the ``magnetic switch'', because it sets in abruptly
when $L_{jet} > L_{crit}$.  Such strong acceleration and collimation near
the rotator should occur only in very special cases when the field is
dynamically dominant and rapidly rotating ({\it e.g.}, in corona above an
accretion disk or rapidly-spinning neutron star or black hole).

\subsection{Attaining Relativistic Speeds:  Strong Magnetic Fields Far from the Central Engine}
\label{sec:attain_hi_speeds}

Some of the microquasar phenomena listed in Sect. \ref{sec:intro} achieve very fast
jet speeds.  Technically, these do not exceed the escape speed from the
surface of a black hole ($v_{jet} = c$), but in practice they seem far too
large to be explained by  the simple MHD wind model discussed in Sect.
\ref{sec:slow_acc_normal}.  How, then, are highly-relativistic speeds attained?
It is unlikely that the magnetic switch process can explain them. 
Not only should this process be rather rare, but observational
evidence discussed in Sect. \ref{sec:slow_acc_normal} rules out the magnetic
switch as the explanation for the high $\Gamma_{jet}$ seen in 
M87 and in many quasars:  those jets seem too broad at their bases
to be rapidly accelerated close to the black hole.  
By analogy, we probably also should look
for another explanation in classical microquasars and GRBs.

One possible model is that of long, straight magnetic fields that
dominate the jet dynamics.  The argument is as follows.  In order to attain
relativistic speeds we need a relativistic Alfven speed, or
\begin{equation}
\Gamma_{A} = V_{A} / c = B / (4 \pi \rho c^{2})^{1/2} >> 1
\label{eq:gamma_alfven}
\end{equation}
Either $B$ must be strong inside the jet or ``mass-loading'' 
of the magnetic field lines must be low.
These requirements present their own problems.  
Any magnetic field carried with the flow has an effective mass density 
of $\rho_{m} \sim B_{\perp}^{2} / 8 \pi c^{2}$, where $B_{\perp}$ is the
component of the field that is perpendicular to the jet flow vector.
If $B_{\perp} \sim B$, then substitution of $\rho_{m}$ into eq.  
(\ref{eq:gamma_alfven}) shows that $\Gamma_{A}$ never will be able to 
attain a value much greater than unity.  One solution to this
problem is to keep the magnetic field well-ordered and straight
over long distances and primarily parallel to the jet flow, so that
$B_{\perp} << B$.  The jet flow then slips along the parallel
magnetic component, which does not contribute to the plasma inertia, only 
occasionally accelerated further by the oscillating and rotating
transverse component $B_{\perp}$.   The acceleration can continue
until $\Gamma_{jet} > \Gamma_{A}$.

Semi-analytic models of such ``Poynting-flux-dominated'' jets have been
built \cite{li92,love02}. However, no numerical simulations of these highly 
relativistic jets have been performed yet.  The best numerical results so 
far are from {\em non}-relativistic simulations \cite{nak01,nak02}, which 
compute  a jet loaded with a decreasing density (increasing $V_{A}$) material.  
In this non-relativistic simulation, the field pitch angle is not so small, 
so $B_{\perp} \sim B_{\parallel}$.  But, the important point is that the flow is 
stable, even over long distances.  The torsional Alfven wave is able to 
transport energy and further accelerate the flow far from its origin. 

\section{MHD Jet-Powered Supernovae:  Microquasars Buried inside Stars}

Several authors have suggested in the past that MHD phenomena may power
supernovae \cite{lw70,bk71}. The recent
discovery that SN ejecta are elongated by an asymmetric jet-like
flow has stimulated renewed interest in these models and, in particular,
in the possibility that an MHD jet produced by the proto-pulsar may
be the source of the explosion energy in all core-collapse SN.
We have proposed \cite{w02} an explosion mechanism that is
consistent with the above properties of MHD jets. The jet
is produced in the iron mantle, just outside the proto-neutron star.
An object with a $10^{15}$ G field and a rotation period of 
$\sim 1$ ms can produce a jet power of $\sim 3 \times 10^{51} {\rm erg~s^{-1}}$ 
and a total energy of $\sim 3 \times 10^{52}$ erg --- more than enough to eject 
the outer envelope and account
for the observed explosion energy.  The proto-neutron star spins down to
more respectable rotation periods ($> 10$ ms) in about $\sim 10$ seconds. 
(The model is similar to that of Ostriker \& Gunn \cite{og71}, with their $10^{12}$ G
pulsar fields replaced by $10^{15}$ G proto-pulsar fields.)  The jet outflow
is composed of iron-rich material and is initially broad at the base.
It therefore can couple well to the outer envelope and eject it.
The flow then collimates at large distances ($> 10^{7-8}$ cm) from the core,
forming a bipolar outflow that imparts an elongated shape to the ejecta.

This model also includes a gamma-ray burst trigger.  In very rare instances, 
the field can be dynamically strong ($B > 10^{16}$ G), leading to satisfaction 
of the magnetic switch condition (eq. \ref{eq:magswitch_jet_power}).
The jet then becomes narrow and very fast, coupling poorly to the mantle, 
punching through the outer envelope \cite{kh01}, escaping the star, and 
producing a heavy iron ``lobe'' outside it, traveling at a speed of 
0.05-0.3 c.  The explosion therefore fails, and much of the mantle falls 
back onto the proto-neutron star, putting the system into a state very 
similar to that at the beginning the ``failed SN'' GRB model \cite{mw99}. 
When the mantle fallback accretes enough material onto the proto-neutron
star (after several minutes to hours), the neutron star is crushed to
a black hole, and a new very fast ($\Gamma_{jet} >> 1$) jet is produced via the
BZ or PC mechanisms discussed above.  The relativistic jet catches up
with the slow, iron-rich lobe at a distance of $d \sim v_{jet} \tau_{fallback}
\sim 10^{12-13}$ cm, and the interaction of jet and lobe produces gamma-rays,
optical afterglow, and an iron-rich spectrum.

\section{Towards a Unified Model for All Galactic Relativistic Jet Sources}

\begin{figure}[htb]
\centering
\psfig{file=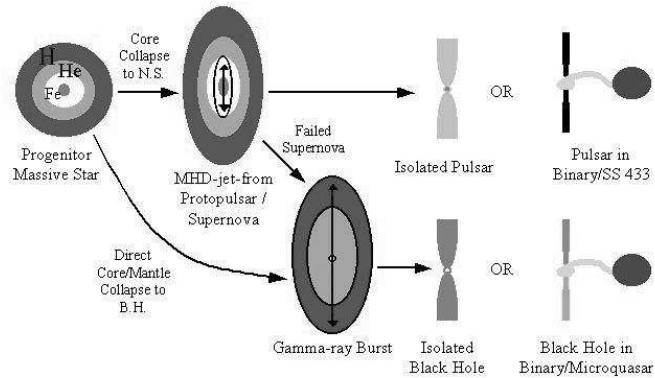,width=10cm}
\caption{The proposed unified model for galactic relativistic jet sources.}
\label{fig:unified_model}
\end{figure}

The possible unification of SN and GRBs, as being different possible
outcomes in the final stages of the death of a massive star and both
caused by MHD-powered jets, strongly suggests a unifying evolutionary
sequence for all galactic sources with relativistic jets.  The model
is shown in Figure \ref{fig:unified_model}, and includes all objects 
discussed in Sect. \ref{sec:intro}.
The sequence begins with a progenitor massive star that is about to
undergo core collapse.  Most collapse to a neutron star, ejecting a broad
MHD jet in the process that drives the SN explosion and produces
the observed asymmetry.  After the envelope dissipates, if the pulsar
is an isolated object, residual rotation of the magnetized remnant
still drives an MHD outflow and a relativistic jet like that seen in
the Crab and Vela pulsars.  If the pulsar resides in a binary system,
it may accrete material from its companion star in a super-Eddington
phase ($\dot{M}_{acc} >> 10^{18} {\rm g~s^{-1}}$) and appear like SS433 for a brief time.
Cessation of the accretion, angular momentum evolution of the pulsar,
and possible collapse to a black hole all could serve to alter SS433's
present state.

In rare circumstances, the MHD SN jet will fail to eject the
envelope, or perhaps the progenitor core will collapse directly to
form a black hole.  In either case a GRB event is generated in a manner
similar to that in the failed SN model.  The GRB jet event will
not spin down the black hole completely (although a significant amount
of rotational energy may be extracted from the black hole by another
means --- gravitational waves).  After the envelope dissipates, if the 
black hole is an isolated object, it will emit little radiation and 
be difficult to detect.  On the other hand, if the newly-formed hole 
is in a binary system, it also can accrete plasma and field from its 
companion, thereby producing a strong jet and classical microquasar. 
Again, changes in the accretion rate and angular momentum evolution of
the black hole will alter the microquasar's observational state. Many 
old microquasars may have masses and angular momenta quite different 
from those they possessed when first formed.

The key element of this unified model is that the final evolutionary
outcome of a massive star core is ultimately determined by the magnitude
and direction of that star core's angular momentum and magnetic field.
We therefore should consider the jets seen in young pulsars and
SS433-type objects to be vestiges of the mechanism that exploded the
massive stars from which they came and, in a similar manner, consider
the jets in microquasars to be the remnants of the gamma-ray burst that
triggered the black hole's formation eons ago.  Jets are the echoes
of violent events of the distant past.  

The $10^{14-15}$ G magnetic 
field strengths needed in SN cores are the real key to the
success of the MHD SN model.  Magnetars are believed to have surface field
strengths of this order, but pulsars typically have fields of order
$10^{12-13}$ G, which are too weak to produce the observed explosion power
or energy.  If stronger fields existed in pulsars at the time they were 
formed, they must have been dissipated either during the SN process 
or shortly thereafter, but it is not known how that dissipation may have 
taken place. Secondly, there also may be competing jet mechanisms (neutrino 
radiation pressure, etc.) which we have not discussed here.
Thirdly, while we have suggested a possible SN failure mechanism
(the magnetic switch), much more detailed theoretical work will be needed
before we will be able to perform the simulations necessary to rule out
different such mechanisms.  Finally, there is a problem
that needs to be addressed by all SN models.  The iron mantle
in the progenitor star is very neutron rich.  If much of it is ejected
(and the $1.4 M_{\odot}$ proto-neutron star is left), then the predicted amount of
{\em r}-process material may be much larger than that observed.  It is a general
problem for all core-collapse SN models to produce a neutron star
remnant while still not over-producing the {\em r}-process elements.

The model also predicts that the formation of a ``microblazars'' \cite{mr99}
will appear as a GRB, along with possible mass sterilizations or extinctions 
on the earth if close enough \cite{th95,sw02}.  A very bright event may
have left a permanent record on one hemisphere of the moon or on
other natural satellites in the solar system.

\section*{Acknowledgments}
This research was performed at the Jet Propulsion Laboratory, California
Institute of Technology, under contract to NASA.

\end{document}